\documentclass[11 pt,oneside,onecolumn,a4paper]{article}
\usepackage{amsmath}
\usepackage{graphicx}
\DeclareGraphicsExtensions{.eps,.ps,.pdf}
\usepackage{bbm}
\usepackage{bm}% bold math
\usepackage{epsfig}
\usepackage[T1]{fontenc}
\usepackage{esint}
\usepackage{amssymb}

\usepackage{lipsum}
mm \leftmargin 5pt \rightmargin 5pt \hoffset= -15mm

\title{Transformation of intermediate 
times in the decays of moving unstable quantum systems via the exponential modes
}
\author{Filippo Giraldi}

\date{\small{School of Chemistry and Physics, University of KwaZulu-Natal\\ 
and National Institute for Theoretical Physics (NITheP)\\
Westville Campus, Durban 4000, South Africa}}

\begin{document}

\maketitle

\def\bbm[#1]{\mbox{\boldmath$#1$}}

\vspace{0em}

%%%%%%%%%%%%%%%%%%%%% Publisher's Area please ignore %%%%%%%%%%%%%%
%\catchline{}{}{}{}{}
%%%%%%%%%%%%%%%%%%%%%%%%%%%%%%%%%%%%%%%%%%%%%%%%%%%%%%%%%%%%%%%%%%%

%\maketitle

%\begin{history}
%%\received{Day Month Year}
%\revised{Day Month Year}
%\accepted{Day Month Year}
%\comby{(xxxxxxxxxx)}
%\end{history}

\vspace{0em}

\begin{abstract}
The transformation of canonical decay laws of moving unstable quantum systems is studied by approximating, over intermediate times, the decay laws at rest with superpositions of exponential modes via the Prony analysis. The survival probability $\mathcal{P}_p(t)$, which is detected in the laboratory reference frame where the unstable system moves with constant linear momentum $p$, is represented by the transformed form $\mathcal{P}_0\left(\varphi_p(t)\right)$ of the survival probability at rest $\mathcal{P}_0(t)$. The transformation of the intermediate times, which is induced by the change of reference frame, is obtained by evaluating the function $\varphi_p(t)$. Under determined conditions, this function grows linearly and the survival probability transforms, approximately, according to a scaling law over an estimated time window. The relativistic dilation of times holds, approximately, over the time window if the mass of resonance of the mass distribution density is considered to be the effective mass at rest of the moving unstable quantum system.
\end{abstract}

\maketitle

\section{Introduction}\label{1}
\vspace{-0em}

The description of the decay laws of unstable systems via the quantum theory has been a central topic of research for decades. The Weisskopf-Wigner decay theory of quantum unstable systems shows that the non-decay or survival probability is approximately exponential over intermediate and long times \cite{WW1,WW2,KF,Khalfin,FondaGirardiRiminiRPP1978}. Considerable effort has been devoted to characterizing the time scale of the exponential regime of the decay. Over very short times the survival probability exhibits vanishing derivative and, consequently, deviates from the exponential law \cite{FondaGirardiRiminiRPP1978}. The duration of this early-time deviation depends on the features of the Hamiltonian operator and on the boundedness of its spectrum \cite{PeresExpDecay1,PeresExpDecay2}. The exponential decay starts early, if compared to the lifetime, in the condition of small coupling \cite{WW1,WW2,ExpDecay1}. Generally, the exponential decay of the survival probability lasts for decades of lifetimes. The Paley Wiener theorem suggests that, over sufficiently long times, the survival probability deviates into slower decays due to the boundedness from below of the Hamiltonian spectrum \cite{PaleyWienerBook}. Usually, the long-time deviations from the exponential decays are determined by the threshold and by the integral and analytical properties of the mass distribution density (MDD). See Refs. \cite{BWKnigth1977,FondaGirardiRiminiRPP1978,Peres1980,MDDlorentzianGislasonPRA1991,NnExpKumarAJP1992,MDDlorentzianJacobovitsAJP1995,SQSnnexpPRA2005,UrbanowskiEPJD2009,UrbanowskiCEJP2009,SDkelkNovak2010,SDkelkNovak2004,NnExpRamKelk2018,GEPJD2015} for details, to name but a few.

In literature, the MDD is usually represented by the Breit-Wigner form, which is a truncated Lorentzian function \cite{BWMDD3,FondaGirardiRiminiRPP1978,MDDlorentzianGislasonPRA1991,KKSJMO1994,MDDlorentzianJacobovitsAJP1995,UrbanowskiEPJD2009,UrbanowskiCEJP2009,BW_Racz_Urb_Xiv2018}. A more general form of MDDs is represented as the product of a function with a simple pole, a threshold factor and a form factor. The form factor has no threshold and no poles and vanishes for high values of the mass variable. The threshold factor results in a power-law profile near the minimum value of the mass (energy) spectrum. The threshold induces the appearance of long-time inverse-power-law decay of the survival probability, while the simple pole of the MDD determines the exponential decay which dominates for several life times \cite{BWKnigth1977,FondaGirardiRiminiRPP1978,UrbanowskiEPJD2009,UrbanowskiCEJP2009,SDkelkNovak2010,SDkelkNovak2004,GEPJD2015}. The long-time inverse-power-law decays have been detected experimentally by measuring the luminescence decays of dissolved organic materials \cite{IPLexperimentPRL2006}.

Fully exponential decay of the survival probability is obtained if the MDD is a Lorentzian over the whole real line \cite{Khalfin}. The MDD is not uniquely determined by the survival probability \cite{KF,MDDlorentzianJacobovitsAJP1995}. In fact, the survival probability is the square modulus of the survival amplitude, and the survival amplitude is the Fourier transform of the MDD. Consequently, the introduction of a phase factor in the survival amplitude which is not linear in time does not alter the exponential decay law. A superposition of exponential decay laws are obtained by MDDs which consist in sums of Breit-Wigner forms \cite{MDDlorentzianJacobovitsAJP1995}. in this case the mass spectrum is considered to be unbounded from below and the contribution of unphysical negative values of the mass spectrum to the survival probability results to be negligible. See Ref. \cite{MDDlorentzianJacobovitsAJP1995} for details. 

Lower and upper cutoff in the mass spectrum and Breit-Wigner forms of MDD have also been considered. If the contribution of the pole spectrum is inside the support, the corresponding survival probability decays exponentially, approximately, over proper times. This time interval is determined by the lower and upper cutoff of the mass spectrum. See Ref. \cite{BWcutoofs1} for details.

Decay laws are often detected in a laboratory reference frame where the unstable quantum system moves with relativistic or ultrarelativistic velocity. For this reason it is essential to understand how the decay laws are transformed by the change of reference frame. This transformation has been evaluated via quantum theory and special relativity in case the unstable system moves with constant linear momentum in the laboratory reference frame \cite{Khalfin,BakamjianPR1961RQT,ExnerPRD1983,HEP_Stef1996,HEP_Shir2004,HEP_Shir2006,TD_StefanovichXivHep2006,UrbPLB2014,GiacosaAPPB2016,GiacosaAPPB2017,UrbAPB2017,GiacosaAHEP2018}. Great effort has been made to interpret the transformation of the decay laws of a moving unstable quantum system in terms of the relativistic dilation of times. As matter of fact, the relativistic dilation of times is found in the transformation of purely exponential decay of the survival probability at rest uniquely if certain conditions hold. In this regard, see Refs. \cite{HEP_Stef1996,HEP_Shir2004,TD_StefanovichXivHep2006,HEP_Shir2006,GiacosaAPPB2016,GiacosaAPPB2017,UrbAPB2017}. The formal transformation of the survival amplitude at rest in the laboratory frame is determined by integral properties of the MDD and by the linear momentum of the moving unstable system. The threshold of the MDD and the linear momentum determine the transformation of the inverse-power-law decays of the survival probability at rest in the laboratory reference frame. Recent analysis has shown that the long-time inverse-power-law decays transform, approximately, according to a scaling relation, by changing reference frame \cite{Gxiv2018}. The scaling factor is determined by the (non-vanishing) lower bound of the mass spectrum and by the linear momentum, and consists in the ratio of the asymptotic value of the instantaneous mass and of the instantaneous mass at rest of the moving unstable system \cite{UrbPLB2014,UrbAPB2017,UrbanowskiEPJD2009,
UrbanowskiCEJP2009,UrbanowskiPRA1994}. The scaling relation reproduces, approximately, the relativistic dilation of times if the lower bound of the mass spectrum is considered to be the effective mass at rest of the moving unstable system. See Ref. \cite{Gxiv2018} for details.

As a continuation of the scenario described above, here, we consider the decay laws at rest over intermediate times. The decay laws are exponential or deviate in slower forms, but are still faster than every inverse power law. In fact, this regime precedes the long time inverse-power-law decay whose transformation is analyzed in Ref. \cite{Gxiv2018}. We intend to evaluate how the decay laws transform over these intermediate times in the laboratory reference frame where the unstable system moves with constant linear momentum. 

The paper is organized as follows. Section \ref{2} is devoted to the description of general moving unstable quantum systems and to the transformation of the decay laws at rest. In Section \ref{3}, the decay laws at rest are approximated with superpositions of exponential modes via the Prony analysis of the modulus of the survival amplitude at rest. In the same Section, the transformed decay laws are obtained in the laboratory reference frame. Section \ref{4} is devoted to the transformation of times which is induced by the change of reference frame and to the appearance of the relativistic dilation of times. Summary and conclusion are reported in Section \ref{5}. Demonstrations of the results are provided in Appendix.

\section{Moving unstable quantum systems}\label{2}

For the sake of clarity, we describe below the decay laws in the laboratory frame $\mathfrak{S}_p$, where the unstable system moves with constant linear momentum $p$, by following Ref. \cite{UrbAPB2017}. Let $\mathcal{H}$ be the Hilbert space of the quantum states of the unstable system. Let the state kets $|m,p\rangle$ be the common eigenstates of the linear momentum $P$ and of the Hamiltonian $H$ self-adjoint operator. The corresponding eigenvalues are respectively $p$, i.e., $P|m,p\rangle =p |m,p\rangle$, and $E(m,p)$, i.e., $H|m,p\rangle =E(m,p) |m,p\rangle$, for every value $m$ which belongs to the continuous spectrum of the Hamiltonian. 
Let $|\phi\rangle$ be the initial state of the unstable quantum system. This state ket belongs to the Hilbert space $\mathcal{H}$ and is represented via the eigenstates $|m,0\rangle$ of the Hamiltonian as follows, $|\phi\rangle=\int_{\mu_0}^{\infty} \langle 0,m||\phi\rangle |m,0\rangle dm$. In this notation, $\langle 0,m|$ and $\langle p,m|$ represent the bras of the state kets $|m,0\rangle$ and $|m,p\rangle$, respectively.

In the rest reference frame $\mathfrak{S}_0$ of the unstable system, the survival amplitude $A_0(t)$ is referred to as the survival amplitude at rest and reads $A_0(t)=\langle \phi| e^{-\imath H t} |\phi\rangle$, where $\imath$ is the imaginary unit. By considering the completeness of the eigenstates of the Hamiltonian, the survival amplitude at rest is expressed via the following integral form \cite{UrbAPB2017,UrbPLB2014,UrbanowskiEPJD2009,UrbanowskiCEJP2009,FondaGirardiRiminiRPP1978},
\begin{eqnarray}
A_0(t)=\int_{\rm{spec}\left\{H\right\} }\omega\left(m\right) 
\exp\left(-\imath m t\right) dm. \label{A0Int}
\end{eqnarray}
The domain of integration $\rm{spec}\left\{H\right\}$ is the continuous spectrum of the Hamiltonian $H$. The function $\omega\left(m\right)$ represents the MDD of the unstable system and is defined as follows, $\omega\left(m\right)=\left|\langle 0,m||\phi\rangle\right|^2$. The MDD is determined by the initial state and by the Hamiltonian of the system via the eigenstates $|m,0\rangle$. At every instant, the non-decay or survival probability is represented by the probability that the unstable system is still in the initially prepared state and, consequently, has not decayed, yet. Therefore, in the rest reference frame of the moving unstable system, the square modulus of the survival amplitude at rest provides the survival probability, $\mathcal{P}_0(t)=\left|A_0(t)\right|^2$, which is referred to as survival probability at rest.

In the laboratory reference frame $\mathfrak{S}_p$ the unstable system is represented by the state ket $|\phi_p\rangle$ which is an eigenstate of the linear momentum $P$ with eigenvalue $p$. The transformed survival amplitude $A_p(t)$ reads $\langle \phi_p|e^{-\imath H t}|\phi_p\rangle$ and is represented by the following form,
\begin{eqnarray}
A_p(t)=\int_{\rm{spec}\left\{H\right\} } \omega\left(m \right)
\exp\left(-\imath \sqrt{ p^2+m^2}t\right) d m.
\label{Aptdef}
\end{eqnarray}
The survival probability, $\mathcal{P}_p(t)$, is provided by the square modulus of the above expression, $\mathcal{P}_p(t)=\left|A_p(t)\right|^2$. The integral expression (\ref{Aptdef}) of the survival probability has been obtained in various ways by adopting quantum theory and special relativity. See 
Ref. \cite{HEP_Stef1996,HEP_Shir2004,HEP_Shir2006,GiacosaAPPB2016,GiacosaAPPB2017,UrbAPB2017,GiacosaAHEP2018} for details.

The survival amplitude $A_p(t)$ has been evaluated analytically in Ref. \cite{HEP_Shir2004} for a Breit-Wigner form $\omega_{BW}\left(m\right)$ of the MDD,
\begin{eqnarray}
\omega_{BW}\left(m\right)= \frac{\Theta\left(m\right)\Gamma }{2 \pi\left(\left(m-M\right)^2+ \Gamma^2/4\right)},
\label{BWMDD}
\end{eqnarray}
where $\Theta\left(m\right)$ represents the Heaviside unit step function, $M$ is the resonance mass of the unstable quantum system and $\Gamma$ is the decay width at rest. The spectrum of the Hamiltonian is the non-negative real line $\left[\right.0,+\infty\left.\right)$. Following Ref. \cite{HEP_Shir2004}, if $\Gamma/M\ll 1$, and either $t>1/\left(10 \Gamma\right)$ or $t\gg1/M$, the survival amplitude is approximated via the Struve function ${\bf H}_1\left(p t\right)$, the Bessel function of first kind $J_1\left(p t\right)$ and the Bessel function of second kind $Y_1\left(p t\right)$ \cite{GradRyzhandBook,PBMISvol1,NISThandbook,AbrSteg},
\begin{eqnarray}
&&\hspace{-4em}A_p(t)\simeq \exp\left(-\imath \sqrt{\left(M-\imath \frac{\Gamma}{2}\right)^2+p^2}t \right)+\frac{\imath \Gamma p}{2 \pi M^2}\Bigg(\frac{\pi}{2}\left({\bf H}_1\left(p t\right)-\imath J_1\left(p t\right)\right)-1 \nonumber \\ &&\hspace{-0.6em}+\left(1+ \imath \frac{p}{M}\right)^{-2}\left(1+\frac{\pi}{2}\left(Y_1\left(p t\right)-
{\bf H}_1\left(p t\right) \right)\right)\Bigg).
\label{ApBWMDD}
\end{eqnarray}
 Over sufficiently long times, $ p t \gg 1$ and either $t>1/\left(10 \Gamma\right)$ or $t\gg 1/M$, the survival amplitude is approximated by a superposition of exponential modes and an inverse power law,
\begin{eqnarray}
&&\hspace{-4em}A_p(t)\simeq
\exp\left(-\left(\left(1+\kappa\right) \frac{\Gamma}{2 \gamma}+\imath \left(1-\kappa\right)M \gamma+ \right)t\right)
\nonumber \\ &&\hspace{-0.6em}+\frac{\Gamma p }{2 M^2\sqrt{2 \pi p t}}\exp\left(\imath \left(pt-\frac{3}{4}\pi\right)\right),
\label{ApBWMDDasympt}
\end{eqnarray}
where $\kappa =\Gamma^2\left(\gamma^2-1\right)/\left(8M^2 \gamma^4\right)$ and 
$$\gamma=\sqrt{1+\frac{p^2}{M^2}}.$$ The parameter $\gamma$ is the relativistic Lorentz factor of a mass at rest which coincides with the mass of resonance $M$ and moves with linear momentum $p$ in the laboratory reference frame $\mathfrak{S}_p$. As $\Gamma/M \ll 1$, the exponential term of the survival amplitude dominates over the inverse power law for various lifetimes \cite{HEP_Shir2004,FondaGirardiRiminiRPP1978} and the survival probability exhibits an approximate exponential decay over these times,
\begin{eqnarray}
\mathcal{P}_p(t)\simeq \exp\left(-\left(1+\kappa\right)\Gamma \frac{t}{ \gamma}\right).
\label{PplongtBWExp1}
\end{eqnarray}
In this exponential-like regime the relativistic dilation of times holds, approximately, for $\kappa\ll 1$. See Ref. \cite{HEP_Shir2004} for details.

Let the MDD exhibit a threshold which is tailored by a nonegative power law near the non-vanishing lower bound  of the mass spectrum $\left[\right. \mu_0, \infty\left.\right)$,
\begin{eqnarray}
\omega\left(m \right)= \left(m-\mu_0\right)^{\alpha}
\omega_0\left(m \right).
\label{Omegaalpha}
\end{eqnarray}
where $\mu_0>0$. The survival amplitude $A_p(t)$ decays as an inverse power law over very long times \cite{Gxiv2018},
\begin{eqnarray}
\hspace{-3em}A_p(t)\simeq \Gamma\left(1+\alpha\right) 
\omega_0\left(\mu_0\right) \exp\left(-\imath \left(\frac{\pi}{2}
\left(1+\alpha\right)+\sqrt{\mu_0^2+p^2} t\right)\right)\left(\frac{\chi_p}{t}\right)^{1+\alpha},
\label{AplongtPLMDD1}
\end{eqnarray}
where
\begin{eqnarray}
\chi_p = \sqrt{1+\frac{p^2}{\mu_0^2}}. \label{Chip}
\end{eqnarray}
Expression (\ref{AplongtPLMDD1}) of the survival amplitude holds for every value of the linear momentum $p$ and provides an inverse-power-law decay of the survival probability over very long times,
 \begin{eqnarray}
P_p(t)\simeq \left(\Gamma\left(1+\alpha\right) 
\omega_0\left(\mu_0\right)\right)^2
\left(\frac{\chi_p}{t}\right)^{2\left(1+\alpha\right)}.
\label{PplongtPLMDD1}
\end{eqnarray}
According to the above expression, the survival probability $\mathcal{P}_p\left(t\right)$ and the survival probability at rest $\mathcal{P}_0\left(t\right)$ are related, approximately over very long times, by the following scaling law,
\begin{eqnarray}
\mathcal{P}_p\left(t\right) \simeq \mathcal{P}_0\left(\frac{t}{ \chi_p}\right). \label{PpP0Lpl1}
\end{eqnarray}
The above relation consists, approximately, in a time dilation with the scaling factor $\chi_p$. This factor is equal to the ratio of the asymptotic value of the instantaneous mass $M_p\left(\infty\right)$ and of the instantaneous mass at rest $M_0\left(\infty\right)$ of the moving unstable system, $\chi_p=M_p\left(\infty\right)/M_0\left(\infty\right)$, where $M_p\left(\infty\right)=\sqrt{\mu_0^2+p^2}$, and 
$M_0\left(\infty\right)=\mu_0$. The scaling factor coincides with the relativistic Lorentz factor of a mass at rest $\mu_0$ which moves with linear momentum $p$ or, equivalently, with constant velocity $1\Big/\sqrt{1+\mu_0^2\big/p^2}$. Consequently, the time dilation of the survival probability reproduces the relativistic time dilation of the instantaneous mass at rest $M_0\left(\infty\right)$ which moves with constant linear momentum $p$. See Ref. \cite{Gxiv2018} for the definition of the expression (\ref{Omegaalpha}) of the MDD, for the estimate of the very long times and for details.

A full exponential decay of the survival probability at rest is obtained by extending the Breit-Wigner form (\ref{BWMDD}) of the MDD to the whole real line. The survival amplitude at rest reads $A_0(t)=\exp \left(-\left(\imath M+\Gamma/2\right)t \right)$ and the survival probability at rest is $\mathcal{P}_0\left(t\right)=\exp \left(-\Gamma t\right)$. Negative values and unboundedness from below of the mass spectrum are obviously unphysical. Still, this approximation properly reproduces the exponential decays which are detected experimentally. The contribution of negative values of the mass spectrum to the survival amplitude is negligible if $\Gamma/M\ll 1$. Following Ref. \cite{GiacosaAPPB2016}, in the laboratory frame $\mathfrak{S}_p$ the survival amplitude is $A_p(t)=\exp \left(-\imath \sqrt{\left(M-\imath \Gamma/2\right)^2+p^2}t\right)$, and the transformed survival probability reads $\mathcal{P}_p\left(t\right)=\exp \left(-\Gamma_p t\right)$. The transformed decay width $\Gamma_p$ reads $\Gamma_p=2 \rm{Im}\left\{\sqrt{\left(M-\imath \Gamma/2\right)^2+p^2}\right\} $. In this case, the scaling factor of the transformed survival probability is $\Gamma/\Gamma_p$. See Refs. \cite{GiacosaAPPB2016,GiacosaAPPB2017,GiacosaAHEP2018} for details.

\section{Decay laws in the laboratory reference frame}\label{3}

At this stage we start our analysis. We consider canonical forms of the survival probability which are monotonically decreasing functions of time. 
In the rest reference frame $\mathfrak{S}_0$ of the moving unstable quantum system we focus on the intermediate regime of the decay. This regime follows the non-exponential decay which appears over very short times \cite{FondaGirardiRiminiRPP1978}. Over the intermediate regime the decay is purely exponential, and includes slower forms, but has not yet become the inverse power law which is determined, over very long times, by the threshold of the MDD \cite{BWKnigth1977,FondaGirardiRiminiRPP1978,UrbanowskiEPJD2009,UrbanowskiCEJP2009,SDkelkNovak2010,SDkelkNovak2004,NnExpRamKelk2018,GEPJD2015}. 
Over these intermediate times we assume the modulus of the survival amplitude at rest to be properly approximated by a superposition of a finite number of exponential modes,
\begin{eqnarray}
\sqrt{\mathcal{P}_0(t)}=\sum_{j=1}^N w_j \exp\left(-\frac{\Gamma_j}{2} t\right). \label{A0ExpSum}
\end{eqnarray}
The above sum is provided by the Prony analysis of the modulus of the survival amplitude at rest. The Prony analysis is a powerful technique of approximation which generalizes the Fourier analysis and estimates the frequency, the damping, the strength and the relative phase of the modal components of a detected signal \cite{Prony1,Prony2,Prony3,Prony4,Prony5}. The parameters $w_1, \ldots, w_N$ represent the normalized weights of the corresponding exponential modes, $\sum_{j=1}^N w_j=1$. The decay widths of the exponential modes are sorted in increasing order, $0<\Gamma_{1}<\ldots<\Gamma_{N}$. It is assumed that the decay widths are negligible with respect to the mass of resonance, $\Gamma_j/M\ll 1$ for every $j=1,\ldots,N$, or, equivalently, $\Gamma_N/M\ll 1$. This property is fundamental for the evaluation of the survival probability in the laboratory reference frame. Expression (\ref{A0ExpSum}) fulfills the following canonical properties of the survival probability of an unstable decaying system, $\mathcal{P}_0(0)=1$ and $\dot{\mathcal{P}}_0(t)<0$ for every $t>0$. The purely exponential decay $\exp\left(-\Gamma_1 t \right)$ of the survival probability at rest $\mathcal{P}_0(t)$ corresponds to the limiting condition where $w_1\to 1^-$ and $w_j\to 0^+$ for every $j=2,\ldots,N$. Refer to \cite{Prony1,Prony2,Prony3,Prony4,Prony5} for a detailed description of the Prony analysis. We stress that expression (\ref{A0ExpSum}) can not approximate the survival probability over very short and very long times. In fact, the survival probability is non-exponential over very short times, as $\dot{\mathcal{P}}_0(0)=0$, and is slower than every exponential decay over very long times \cite{Khalfin,FondaGirardiRiminiRPP1978,PaleyWienerBook}.

We intend to evaluate the survival probability $\mathcal{P}_p(t)$ in the laboratory reference frame $\mathfrak{S}_p$, where the unstable system moves with constant linear momentum $p$. The following assumptions motivate the choice of approximating the modulus of the survival amplitude at rest by the superposition (\ref{A0ExpSum}) of exponential modes. Over intermediate times, the decay laws are mainly determined by the behavior of the MDD around the mass of resonance $M$. The MDD is approximately symmetric with respect to the mass of resonance, 
\begin{eqnarray}
\omega \left(M+m^{\prime}\right)=\omega \left(M-m^{\prime}\right). \label{symmMDD}
\end{eqnarray}
These assumptions are fundamental for the present analysis and are based upon the following observations \cite{MDDlorentzianGislasonPRA1991,NnExpKumarAJP1992,MDDlorentzianJacobovitsAJP1995,GPoscillatingDecaysQM2012,GiacosaAPPB2016,GiacosaAPPB2017}. Purely exponential decays of the survival probability are obtained from Lorentzian MDDs through the complex-valued simple pole of the Lorentzian function. Lorentzian MDDs are symmetric with respect to the mass of resonance. The support of these MDDs is the whole real line and the contribution of the unphysical negative values of the mass spectrum to the survival amplitude is negligible. This condition is satisfied as $\Gamma_N/M\ll 1$. 
. 

The survival probability $\mathcal{P}_p(t)$, which is detected in the laboratory reference frame $\mathfrak{S}_p$, is evaluated as the square modulus of the survival amplitude $A_p(t)$, which is given by Eq. (\ref{Aptdef}). The fundamental assumptions about the symmetry of the MDD and about the negligible contribution of the negative values of the mass spectrum lead to the following 
relation,
\begin{eqnarray}
\hspace{-0em}\mathcal{P}_p(t)=\left|\frac{2}{\pi}\int_0^{\infty}
\exp\left(-\imath \sqrt{p^2+m^2}t\right)
dm \int_0^{\infty}\sqrt{\mathcal{P}_0\left(t^{\prime}\right)} \cos\left(M t^{\prime}\right)\cos\left(m t^{\prime}\right) dt^{\prime} \right|^2, \label{PpP0Int1}
\end{eqnarray}
which links the survival probability $\mathcal{P}_p(t)$ and the survival probability at rest $\mathcal{P}_0(t)$. If the modulus of the survival amplitude at rest is given by Eq. (\ref{A0ExpSum}), relation (\ref{PpP0Int1}) provides the following expression,
\begin{eqnarray}
&&\hspace{-3em}\mathcal{P}_p(t)\simeq\Bigg|\sum_{j=1}^N
w_j \exp\left(-\frac{1}{2}\Upsilon\left(M,\Gamma_j,p\right)t\right) 
+\imath \frac{p \sum_{j=1}^N
w_j\Gamma_j}{\pi M^2} \Xi\left(M,p,t\right)\Bigg|^2. \label{PptHYJ}
\end{eqnarray} 
This form approximates the survival probability $\mathcal{P}_p(t)$ in the laboratory reference frame $\mathfrak{S}_p$, over times $t$ such that either $t>1/\left(10 \Gamma_1\right)$ or $M t \gg 1$. These conditions and expression (\ref{PptHYJ}) are obtained from the analysis of the survival amplitude $A_p(t)$ which is performed in Ref. \cite{HEP_Shir2004}. Some details of that analysis are reported in the forth paragraph of Section \ref{2}, for the sake of clarity. Notice that expression (\ref{PptHYJ}) does not describe the survival probability over very short times, $ 1 \geq 10 \Gamma_1 t$ and $1 \gtrsim M t$. The function $\Upsilon\left(M,\Gamma,p\right)$ is defined, for every value of the mass of resonance $M$, of the decay width $\Gamma$ and of the linear momentum $p$, by the form below \cite{HEP_Shir2004},
\begin{eqnarray}
\Upsilon\left(M,\Gamma,p\right)=\Lambda_-\left(M,\Gamma,p\right)
 + \imath \Lambda_+\left(M,\Gamma,p\right), \label{Upsilon}
\end{eqnarray}
where 
\begin{eqnarray}
&&\hspace{-3em}\Lambda_{\mp}\left(M,\Gamma,p\right)=\sqrt{2\left(\sqrt{\left(M^2-\frac{\Gamma^2}{4}+p^2\right)^2+M^2 \Gamma^2}\mp\left(M^2-\frac{\Gamma^2}{4}+p^2\right)\right)}. \label{Lambdamp}
\end{eqnarray}
The function $\Xi\left(M,p,t\right)$ is defined, for every value of the mass of resonance $M$ and of the linear momentum $p$, via the Bessel Function $J_1\left(pt\right)$, the modified Bessel function $Y_1\left(pt\right)$
and the Struve function $\mathbf{H}_1\left(pt\right)$ as below \cite{HEP_Shir2004},
\begin{eqnarray}
&&\hspace{-2.2em}\Xi\left(M,p,t\right)=\frac{\pi }{2}\left(\mathbf{H}_1\left(pt\right)-\imath J_1\left(pt\right)\right)-1+
\frac{1-p^2/M^2}{\left(1+p^2/M^2\right)^2} \nonumber \\ &&\hspace{3.2em}\times \Big(1+\frac{\pi }{2}\left(Y_1\left(pt\right)-\mathbf{H}_1\left(pt\right)\right)\Big). %\nonumber
\label{Xi}
\end{eqnarray}
Refer to \cite{NISThandbook,GradRyzhandBook,AbrSteg} for details on the asymptotic properties of these special functions.

\subsection{Exponential times in the laboratory reference frame}\label{31}

In the laboratory reference frame $\mathfrak{S}_p$ the survival probability, which is described by Eq. (\ref{PptHYJ}), results, approximately, in a superposition of exponential modes (exponential-like decay) over intermediate times, and becomes, approximately, an inverse power law over very long times \cite{HEP_Shir2004}. We intend to estimate the intermediate times in the laboratory reference frame $\mathfrak{S}_p$ via the mass of resonance $M$, the decay widths $\Gamma_1,\ldots,\Gamma_N$ of the exponential modes at rest, and the linear momentum $p$. Here, we refer to these times as the exponential times \cite{HEP_Shir2004}.

As $\Gamma_N/M \ll 1$, the first sum appearing in the right side of Eq. (\ref{PptHYJ}) is approximated by the following form \cite{HEP_Shir2004},
\begin{eqnarray}
\hspace{-2em}
\sum_{j=1}^N w_j \exp\left(- \frac{1}{2}\Upsilon\left(M,\Gamma_j,p\right)t\right)\simeq \sum_{j=1}^N w_j \exp\left(- \frac{t}{2}\left(\frac{\Gamma_j}{\gamma}+2 \imath M \gamma\right)\right). \label{PpExp1}
\end{eqnarray}
Instead, for $p t \gg 1$, the function $\Xi\left(M,p,t\right)$ is approximated as below,
\begin{eqnarray}
\Xi\left(M,p,t\right) \simeq - \imath \sqrt{ \frac{\pi}{2 p t}} \exp\left(\imath\left( pt - \frac{3}{4}\pi\right)\right). \label{XipPl1}
\end{eqnarray}
Consequently, the modulus of the first term appearing in the right side of Eq. (\ref{PptHYJ}) is a superposition of the transformed exponential modes, while the modulus of the second term is proportional to the inverse power law $1/\sqrt{pt}$ over times $t$ such that $p t \gg 1$, and either $t>1/\left(10 \Gamma_1\right)$ or $M t \gg 1$. Therefore, we evaluate the exponential times as those times over which one, or more, transformed exponential mode dominates over the inverse-power-law term, and such that $p t \gg 1$, and either $t>1/\left(10 \Gamma_1\right)$ or $M t \gg 1$. Over these times the survival probability, given by Eq. (\ref{PptHYJ}), is approximated by the square modulus of the sum of the selected exponential modes. The resulting decay is, approximately, exponential-like in the laboratory reference frame $\mathfrak{S}_p$.

The exponential times and the dominant exponential modes are found via the following technique. Let the parameter $\xi_j$ be defined as below,
\begin{eqnarray}
\xi_j=\frac{\sum_{i=1}^N w_i \Gamma_i/w_j}{2 M} 
\sqrt{\frac{\Gamma_j }{\pi  M}\sqrt{1-\frac{1}
{\gamma^2}}}, \label{xij}
\end{eqnarray}
for every $j=1,\ldots,N$. The indexes $j_1,\ldots,j_{n_0}$ are chosen among the indexes $1,\ldots,N$, in such a way that the constraint $\xi_{j_l} \ll 10^{-2}$ is fulfilled for every $l=1,\ldots,n_0$. These indexes are sorted in increasing order, $j_1<\ldots<j_{n_0}$, with $1\leq n_0 \leq N$. For example, let the order of magnitude of the quantity $\sum_{i=1}^N w_i \Gamma_i/M$ and of the ratio $\Gamma_j/M$ be less than or equal to $\left(-4\right)$. Let the order of magnitude of the weight $w_j$ be equal to $\left(-1\right)$. Under these conditions the constraint $\xi_j\ll 10^{-2}$ is fulfilled. For the sake of convenience we define the closed time interval $I_{p,l}$, in the laboratory reference frame $\mathfrak{S}_p$, and the closed time interval $I_{0,l}$, in the rest reference frame $\mathfrak{S}_0$, as below,
\begin{eqnarray}
I_{p,l}=\left[\frac{2\zeta_{\rm min}\gamma}{\Gamma_{j_l}},\frac{2\zeta_{\rm max}\gamma}{\Gamma_{j_l}}\right], \hspace{1em} 
I_{0,l}=\left[\frac{2\zeta_{\rm min}}{\Gamma_{j_l}},\frac{2\zeta_{\rm max}}{\Gamma_{j_l}}\right],
\label{IpI0l}
\end{eqnarray}
for every $l=1,\ldots,n_0$. The parameters $\zeta_{\rm min}$ and $\zeta_{\rm max}$ are defined in Appendix and are approximated by the following values, $\zeta_{\rm min}\simeq 0.0001$ and $\zeta_{\rm max}\simeq 5.4533$. We are finally equipped to estimate the exponential times in terms of the decay widths $\Gamma_1,\ldots,\Gamma_N$, the mass of resonance $M$ and the relativistic Lorentz factor $\gamma$. The exponential times are approximated by the set of times $I_p$ which is defined as the union of the time intervals $I_{p,1}, \ldots, 
I_{p,n_0}$,
\begin{eqnarray}
I_{p}=\bigcup_{l=1}^{n_0}I_{p,l}.
\label{Ip}
\end{eqnarray}
The set $I_p$ is a closed interval if $n_0=1$ and in other cases. If $n_0>1$, let the inequality 
\begin{eqnarray}
\frac{\Gamma_{j_l}}{\Gamma_{j_{l+1}}}> \frac{\zeta_{\rm min}}{\zeta_{\rm max}}\simeq 1.83 \times 10^{-5}, \label{cond1windows}
\end{eqnarray}
hold for every $l=1,\ldots,n_0-1$. Then, in the laboratory reference frame $\mathfrak{S}_p$ the set of exponential times $I_p$ coincides with the following closed interval,
\begin{eqnarray}
I_p=\left[\frac{2 \zeta_{\rm min} }{\Gamma_{j_{n_0}}}\gamma,\frac{2 \zeta_{\rm max} }{\Gamma_{j_1}}\gamma\right]. \label{Ipclosedint1}
\end{eqnarray}
In general, if the set $I_p$ is a closed interval, then, it is given by Eq. (\ref{Ipclosedint1}). The present approach provides the time window
\begin{eqnarray}
\frac{2 \zeta_{\rm max} }{\Gamma_{j_1}}\gamma\gtrsim t\gtrsim\frac{2 \zeta_{\rm min} }{\Gamma_{j_{n_0}}}\gamma, \label{ExpTn0}
\end{eqnarray}
as an approximate estimate of the exponential times in the laboratory reference frame $\mathfrak{S}_p$.

The exponential times are required to fulfill the constraints under which the inverse-power-law behavior of the transformed survival probability appears. Consequently, the following constraints must hold over the set $I_p$ of the exponential times, $p t \gg 1$, and either $t>1/\left(10 \Gamma_1\right)$ or $M t \gg 1$. The constraint $t>1/\left(10 \Gamma_1\right)$ is fulfilled over the set $I_p$ if $20 \zeta_{\rm min}\gamma>\Gamma_{j_{n_0}}/\Gamma_1$. In case this constraint is not realized, the condition $M t \gg 1$ holds over the set $I_p$ if $2 \zeta_{\rm min}\gamma\gg\Gamma_{j_{n_0}}/M$. The condition $p t \gg 1$ holds over the set $I_p$ if $2 \zeta_{\rm min}\gamma \sqrt{\gamma^2-1}\gg\Gamma_{j_{n_0}}/M$. The nonrelativistic limit, $\gamma\to 1^+$, is excluded. In the decays of unstable particles, the values of the ratio between the decay width $\Gamma$ and the mass of resonance $M$ can be quite small. For example, this ratio is $\Gamma/M\simeq 2.83 \times 10^{-18}$ in the decay of the muon particle, and reads $\Gamma/M\simeq 5.72 \times 10^{-8}$ in the decay of the neutral pion, while it increases for strong decays, $\Gamma/M\simeq 0.191$ in the decay of the $\rho$ meson. See Refs. \cite{DecayData,GiacosaAPPB2016} for details. Qualitatively, the value $\Gamma_{j_{n_0}}$, which is provided by the present approach, is expected to be close to the decay width $\Gamma$ nearby the exponential regime of the decay. Except for strong decays and for the nonrelativistic regime, the above constraints are expected to hold over the exponential times. If these constraints do not hold, the method which is reported in the previous paragraph, and explained in Appendix, can be adjusted by choosing a different order of magnitude for the parameter $\xi_j$ and by changing the values of the parameters $\zeta_{\rm min}$ and $\zeta_{\rm max}$, accordingly. See Appendix for details.

At this stage, we describe the decay laws in the laboratory reference frame $\mathfrak{S}_p$ over the exponential times. The survival probability $\mathcal{P}_p(t)$ is properly approximated by the following exponential-like decay, 
\begin{eqnarray}
\hspace{-2em}
\mathcal{P}_p(t) \simeq \left| \sum_{l} {}^{'} w_{j_l} \exp\left(- \frac{\Gamma_{j_l} t}{2\gamma}\right)\right|^2, \label{ExpDecay1n0}
\end{eqnarray}
for every instant of the exponential times, $ \forall \,\,t\in I_p$. The sum $\sum_{l} {}^{'}$ depends on the time $t$ as follows. Every index $l$, over which the $\sum_{l} {}^{'}$ is performed, is chosen among the indexes $1,\ldots,n_0$, in such a way that $t \in I_{p,l}$. % for every instant $t$ of the exponential times. 
At every instant of the set $I_p$ of the exponential times, the exponential modes which are selected for the sum $\sum_{l} {}^{'}$ dominate over the remaining exponential terms, which appear in Eq. (\ref{PpExp1}), and over the inverse-power-law term. See Appendix for detail. By substituting the ratio $t/\gamma$ with $t$ in Eq. (\ref{ExpDecay1n0}), the present technique selects the dominant terms among the exponential modes of Eq. (\ref{A0ExpSum}) for every $t \in I_0$. This set of times $I_0$ is defined in the rest reference frame $\mathfrak{S}_0$ as below,
\begin{eqnarray}
I_{0}=\bigcup_{l=1}^{n_0}I_{0,l}.
\label{I0}
\end{eqnarray} 
Consequently, in the rest reference frame $\mathfrak{S}_0$, the survival probability $\mathcal{P}_0(t)$ is properly approximated by the form
\begin{eqnarray}
\hspace{-2em}
\mathcal{P}_0(t) \simeq \left| \sum_{l} {}^{''} w_{j_l} \exp\left(- \frac{\Gamma_{j_l} t}{2}\right)\right|^2, \label{ExpDecay1n0approxRest}
\end{eqnarray}
over the set of times $I_0$.  The sum $\sum_{l} {}^{''}$ depends on the time $t$ as follows. Every index $l$, over which the $\sum_{l} {}^{''}$ is performed, is chosen among the indexes $1,\ldots,n_0$, in such a way that $t \in I_{0,l}$. If $n_0=1$, or if $n_0>1$ and condition (\ref{cond1windows}) holds, the set $I_0$ consists in the following closed time interval,
\begin{eqnarray}
I_0=\left[\frac{2 \zeta_{\rm min} }{\Gamma_{j_{n_0}}},\frac{2 \zeta_{\rm max} }{\Gamma_{j_1}}\right]. \label{I0window}
\end{eqnarray}
This property is crucial and is used in Section \ref{4} for the study of the transformation of times which is due to the change of 
reference frame.

In summary, starting from the Prony analysis (\ref{A0ExpSum}) of the modulus of the survival amplitude at rest, we have evaluated the transformed survival probability, Eq. (\ref{PptHYJ}), in the laboratory reference frame $\mathfrak{S}_p$, and the time window (\ref{ExpTn0}) over which the survival probability is approximately exponential-like, Eq. (\ref{ExpDecay1n0}). This description is the first of the two main results of the paper.

\subsection{Transformation of stretched exponential decays via the Prony analysis}\label{32}

Over intermediate times the purely exponential forms of canonical decays are supposed to evolve in slower forms which end up in inverse power laws over very long times. Stretched exponential functions are arbitrarily slower than every purely exponential decay and are faster than every inverse power law. Therefore, we choose to test the present theoretical construct and find the transformed forms of stretched exponential decays, at rest, in the laboratory reference frame $\mathfrak{S}_p$.
 We stress that this choice is not supported by any experimental work. However, this test allows to observe, numerically, how small deviations from exponential decays, which occur in the rest reference frame $\mathfrak{S}_0$, are detected in the laboratory reference frame $\mathfrak{S}_p$, according to the present theoretical approach. 

As matter of fact, the stretched exponential decay \cite{StretchExp1,StretchExp2}, 
\begin{eqnarray}
\mathcal{P}_0(t)\simeq
\exp\left(-\left(\frac{t}{\overline{\tau}}\right)^{\vartheta}\right), 
\label{P0stretched1def}
\end{eqnarray}
is the natural generalization of the purely exponential decay, which is found for $\vartheta\to 1^-$. The positive power $\vartheta$ is smaller than unity, $0<\vartheta<1$, while the parameter $\overline{\tau}$ represents the characteristic time of the stretched exponential decay. The transformation of the decay law (\ref{P0stretched1def}) in the laboratory reference frame $\mathfrak{S}_p$ is obtained from Eqs. (\ref{A0ExpSum})-(\ref{PptHYJ}), by performing the Prony analysis \cite{Prony1,Prony2,Prony3,Prony4} of the stretched exponential decay, 
\begin{eqnarray}
\exp\left(-\frac{1}{2}\left(\frac{t}{\overline{\tau}}\right)^{\vartheta}\right)\simeq \sum_{j=1}^{\overline{N}} \overline{w}_j \exp\left(- \frac{\overline{\Gamma}_j }{2}t\right). 
\label{P0stretched1}
\end{eqnarray} 
Following Ref. \cite{Prony4}, the values of the coefficients $\overline{w}_1, \ldots$, $\overline{w}_{\overline{N}}$, and $\overline{\Gamma}_1, \ldots$, $\overline{\Gamma}_{\overline{N}}$, are initially selected via a Monte Carlo approach and then chosen by minimizing the root-mean-square error. In Ref. \cite{Prony4} the computation is performed for $\vartheta=3/5,1/2,3/7$, and for different values of the number $\overline{N}$. Except for small values of the dimensionless quantity $t\big/\left(2^{1/\vartheta}\overline{\tau}\right)$, the finite sum of exponential modes properly reproduces the stretched exponential decays even for small values of the number $N$. See Ref. \cite{Prony4} for details. 

\begin{figure*}
 \includegraphics[width=0.5\textwidth]{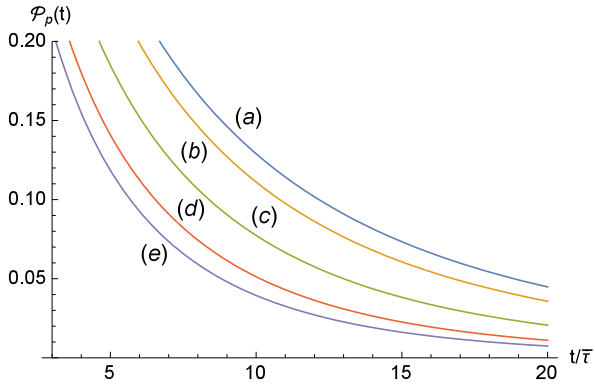}
\caption{(Color online) Transformed survival probability $\mathcal{P}_p(t)$ versus $ t/\overline{\tau}$, for $ 3 \leq t/\overline{\tau}  \leq 20$, and different values of the quantities $p\overline{\tau}$ and $M\overline{\tau}$ and of the Lorentz factor $\gamma$. The corresponding survival probability at rest $\mathcal{P}_0(t)$ is described via Eqs. (\ref{P0stretched1def}) and (\ref{P0stretched1}) with $\vartheta=3/5$. Curve $(a)$ corresponds to $p\overline{\tau}=2000$, $M\overline{\tau}=700$ and $\gamma\simeq 3.0271$. Curve $(b)$ corresponds to $p\overline{\tau}=1500$, $M\overline{\tau}=600$ and $\gamma\simeq 2.6926$. Curve $(c)$ corresponds to $p\overline{\tau}=1100$, $M\overline{\tau}=600$ and $\gamma=2.0883$. Curve $(d)$ corresponds to $p\overline{\tau}=900$, $M\overline{\tau}=700$ and $\gamma=1.6288$. Curve $(e)$ corresponds to $p\overline{\tau}=600$, $M\overline{\tau}=600$ and $\gamma=\sqrt{2}$. }\label{fig1}
\end{figure*}

\begin{figure*}
 \includegraphics[width=0.5\textwidth]{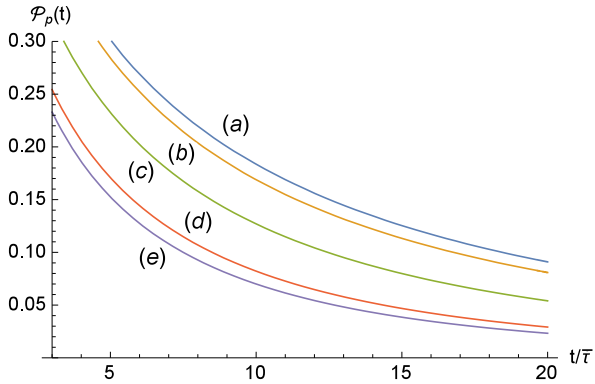}
\caption{(Color online) Transformed survival probability $\mathcal{P}_p(t)$ versus $ t/\overline{\tau}$, for $ 3 \leq t/\overline{\tau}  \leq 20$, and different values of the quantities $p\overline{\tau}$ and $M\overline{\tau}$ and of the Lorentz factor $\gamma$. The survival probability at rest $\mathcal{P}_0(t)$ is described via Eqs. (\ref{P0stretched1def}) and (\ref{P0stretched1}) with $\vartheta=1/2$. Curve $(a)$ corresponds to $p\overline{\tau}=2000$, $M\overline{\tau}=600$ and $\gamma\simeq 3.4801$. Curve $(b)$ corresponds to $p\overline{\tau}=1800$, $M\overline{\tau}=600$ and $\gamma=\sqrt{10}$. Curve $(c)$ corresponds to $p\overline{\tau}=1700$, $M\overline{\tau}=800$ and $\gamma \simeq 2.3485$. Curve $(d)$ corresponds to $p\overline{\tau}=1000$, $M\overline{\tau}=800$ and $\gamma\simeq 1.601$. Curve $(e)$ corresponds to $p\overline{\tau}=700$, $M\overline{\tau}=700$ and $\gamma=\sqrt{2}$.}\label{fig2}
\end{figure*}

Numerical analysis of the survival probability $\mathcal{P}_p(t)$ is displayed in Figures \ref{fig1} and \ref{fig2} in case the survival probability at rest $\mathcal{P}_0(t)$ is given by the stretched exponential law (\ref{P0stretched1def}) and is approximated by the finite sum of exponential modes (\ref{P0stretched1}). The transformed survival probability $\mathcal{P}_p(t)$ is evaluated via Eqs. (\ref{PptHYJ}) and (\ref{P0stretched1}). The values of the parameters $\overline{w}_1, \ldots$, $\overline{w}_{\overline{N}}$, and $\overline{\Gamma}_1, \ldots$, $\overline{\Gamma}_{\overline{N}}$, are derived from the Prony analysis of the stretched exponential decay with $\vartheta=3/5$ and $\vartheta=1/2$, which is performed in Ref. \cite{Prony4}.

\section{Transformation of intermediate times and relativistic time dilation}\label{4}

We intend to study the general transformation of times which occurs by observing the decays of moving unstable quantum systems in the laboratory reference frame $\mathfrak{S}_p$. This transformation is described by the function $\varphi_p(t)$ which is defined with the following relation,
\begin{eqnarray}
\mathcal{P}_p(t)=\mathcal{P}_0\left(\varphi_p(t)\right). \label{PpP0def}
\end{eqnarray}
The survival probability at rest $\mathcal{P}_0\left(t\right)$ is monotonically decreasing for the canonical decay laws under study , $\dot{\mathcal{P}}_0\left(t\right)<0$ for every $t>0$. Consequently, the inverse function $\mathcal{P}^{-1}_0(r)$ exists and is properly defined over the interval $\left(0\right.,\left.1\right]$. This means that the following relations, $\mathcal{P}^{-1}_0\left(\mathcal{P}_0(t)\right)=\mathcal{P}_0\left(\mathcal{P}^{-1}_0(t)\right)=t$, hold for every $t\geq 0$. An analytical description of this inverse function, $\mathcal{P}^{-1}_0: \left(0\right.,1\left.\right]\to \left[0\right.,+\infty\left.\right)$, is provided in Appendix by considering the Prony analysis of the modulus of the survival amplitude at rest, Eq. (\ref{A0ExpSum}). Definition (\ref{PpP0def}) provides the formal expression of the function $\varphi_p(t)$, 
\begin{eqnarray}
\varphi_p(t)=\mathcal{P}^{-1}_0\left(\mathcal{P}_p(t)\right), \label{phipP0Pp}
\end{eqnarray}
for every $t \geq 0$. Once the inverse function $\mathcal{P}^{-1}_0\left(r\right)$ is evaluated, numerically, or via Eqs. (\ref{P0invr}) and (\ref{EqXir}), an analytical description of the function $\varphi_p(t)$ is obtained by approximating the survival probability $\mathcal{P}_p(t)$ via relation (\ref{PptHYJ}),
\begin{eqnarray}
&& \hspace{-4.8em}\varphi_p(t)\simeq \mathcal{P}^{-1}_0\Bigg(\Bigg|\sum_{j=1}^N
w_j \exp\left(- \frac{1}{2}\Upsilon\left(M,\Gamma_j,p\right)t\right) 
+\imath \frac{p \sum_{j=1}^N
w_j\Gamma_j}{\pi M^2} \Xi\left(M,p,t\right)\Bigg|^2\Bigg). \label{phipP0PpHYJ}
\end{eqnarray}
We remind that the above form holds if the condition $\Gamma_N/M \ll 1$ is satisfied, and over times $t$ such that either $t>1/\left(10 \Gamma_1\right)$ or $M t \gg 1$.

The comparison between Eq. (\ref{ExpDecay1n0}) and Eq. (\ref{ExpDecay1n0approxRest}) suggests that, over the set $I_p$ of the exponential times, the survival probability $\mathcal{P}_p(t)$ is approximately related to the survival probability at rest $\mathcal{P}_0 \left(t\right)$ by the following scaling law, 
\begin{eqnarray}
\mathcal{P}_p(t)\simeq \mathcal{P}_0 \left(\frac{t}{\gamma}\right). \label{PpPprel}
\end{eqnarray}
The above expression is classified, here, as an approximation since Eq. (\ref{ExpDecay1n0}) and Eq. (\ref{ExpDecay1n0approxRest}) are, effectively, approximations of the survival probability $\mathcal{P}_p(t)$ and of the survival probability at rest $\mathcal{P}_0(t)$, respectively. The scaling law (\ref{PpPprel}) reproduces, approximately, the relativistic dilation of times if the mass of resonance $M$ is considered to be the effective mass at rest of the unstable quantum systems which moves with constant linear momentum $p$.
Furthermore, the scaling law (\ref{PpPprel}) suggests that the function $\varphi_p(t)$ is approximately linear over the set $I_p$ of the exponential times,
\begin{eqnarray}
\varphi_p(t) \simeq \frac{t}{\gamma}, \label{varphilinear}
\end{eqnarray}
for every $t \in I_p$. If condition (\ref{cond1windows}) holds, the linear growth appears, approximately, over the time window (\ref{ExpTn0}). Equivalently, the survival probability at rest $\mathcal{P}_0(t)$ transforms in the set of times $I_0$, in the rest reference frame $\mathfrak{S}_0$, according to the relativistic dilation of times.  Consequently, in the rest reference frame $\mathfrak{S}_0$ the relativistic dilation of times holds, approximately, over the time window
\begin{eqnarray}
\frac{2 \zeta_{\rm max} }{\Gamma_{j_1}}\gtrsim t\gtrsim\frac{2 \zeta_{\rm min} }{\Gamma_{j_{n_0}}}. \label{ExpTn0p0}
\end{eqnarray} 
Notice that the length $T_{0}$ of the time interval $I_{0}$ is given by $T_{0}=2 \left(\zeta_{\rm max}/\Gamma_{j_1}-\zeta_{\rm min}/\Gamma_{j_{n_0}}\right)$ and transforms in the length $T_{p}$ of the time interval $I_{p}$ according to the relativistic time dilation, $T_{p}= \gamma T_{0}$, .

In summary, we have described the transformation of times in the decay laws of moving unstable quantum systems, which are induced by the change of reference frame, via relations (\ref{PpP0def})-(\ref{phipP0PpHYJ}), and we have found that, under determined conditions, the relativistic dilation of times, Eqs. (\ref{PpPprel}) and (\ref{varphilinear}), holds, approximately, over the time window (\ref{ExpTn0}), in the laboratory reference frame, and over the time window (\ref{ExpTn0p0}), in rest reference frame. These descriptions and properties constitute the last of the two main results of the paper.

\begin{figure*}
 \includegraphics[width=0.5\textwidth]{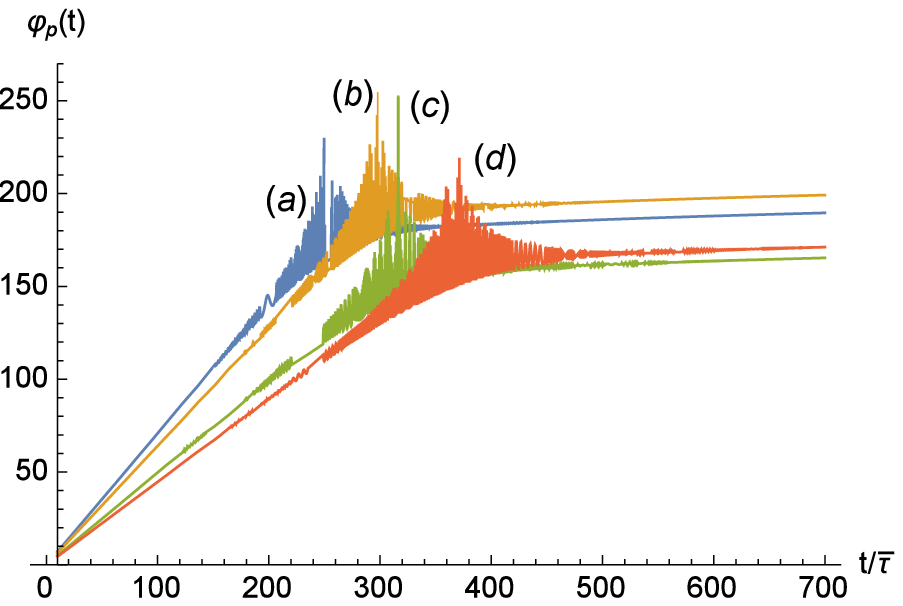}
\caption{(Color online) Function $\varphi_p(t)$ versus $ t/\overline{\tau}$, for $ 10 \leq t/\overline{\tau}  \leq 700$, and different values of the quantities $p\overline{\tau}$ and $M\overline{\tau}$ and of the corresponding Lorentz factor $\gamma$. The survival probability at rest $\mathcal{P}_0(t)$ is descried via Eqs. (\ref{P0stretched1def}) and (\ref{P0stretched1}) with $\vartheta=3/5$. Curve $(a)$ corresponds to $p\overline{\tau}=700$, $M\overline{\tau}=700$ and $\gamma=\sqrt{2}$. Curve $(b)$ corresponds to $p\overline{\tau}=1200$, $M\overline{\tau}=1000$ and $\gamma\simeq 1.5620$. Curve $(c)$ corresponds to $p\overline{\tau}=700$, $M\overline{\tau}=400$ and$\gamma\simeq 2.0156$. Curve $(d)$ corresponds to $p\overline{\tau}=1000$, $M\overline{\tau}=500$ and $\gamma =\sqrt{5}$.}\label{fig3}
\end{figure*}

\begin{figure*}
 \includegraphics[width=0.5\textwidth]{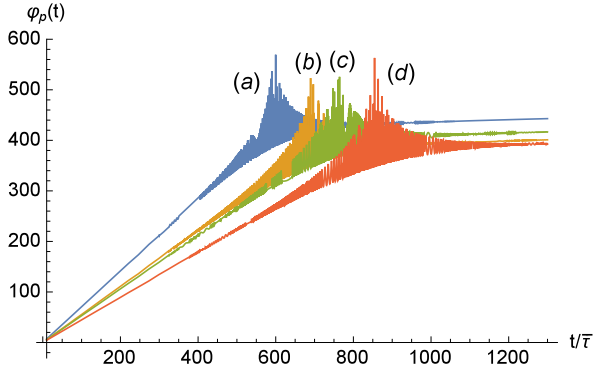}
\caption{(Color online) Function $\varphi_p(t)$ versus $ t/\overline{\tau}$, for $ 10 \leq t/\overline{\tau}  \leq 1300$, and different values of the quantities $p\overline{\tau}$ and $M\overline{\tau}$ and of the corresponding Lorentz factor $\gamma$. The survival probability at rest $\mathcal{P}_0(t)$ is described via Eqs. (\ref{P0stretched1def}) and (\ref{P0stretched1}) with $\vartheta=1/2$. Curve $(a)$ corresponds to $p\overline{\tau}=600$, $M\overline{\tau}=600$ and $\gamma=\sqrt{2}$. Curve $(b)$ corresponds to $p\overline{\tau}=600$ and $M\overline{\tau}=400$ and $\gamma\simeq 1.8028$. Curve $(c)$ corresponds to $p\overline{\tau}=800$, $M\overline{\tau}=500$ and $\gamma \simeq 1.8868$. Curve $(d)$ corresponds to $p\overline{\tau}=800$, $M\overline{\tau}=400$ and $\gamma\simeq 2.2361$.}\label{fig4}
\end{figure*}

Numerical analysis of the function $\varphi_p(t)$ is displayed in Figures \ref{fig3} and \ref{fig4}. The survival probability at rest $\mathcal{P}_0(t)$ corresponds to the 
stretched exponential form (\ref{P0stretched1def}) with $\vartheta =3/5$, Figure \ref{fig3}, and $\vartheta =1/2$, Figure \ref{fig4}. The computation is performed by approximating the survival probability at rest via the Prony analysis, Eq. (\ref{P0stretched1}), of the stretched exponential decay which is reported in Ref. \cite{Prony4}
 for $\vartheta=3/5$ and $\vartheta=1/2$ and for $\overline{N}=8$. The transformed survival probability $\mathcal{P}_p(t)$ is evaluated via Eq. (\ref{PptHYJ}). The linear growth of the curves, which is displayed in Figures \ref{3} and \ref{4}, agrees with the scaling relation (\ref{varphilinear}) which holds, approximately, over the exponential time window (\ref{ExpTn0}).

\section{Summary and conclusions}\label{5}

In many high-energy accelerator experiments and astrophysical phenomena the decay processes are observed in laboratory reference frames where unstable particles move at relativistic or 
ultrarelativistic velocities. The corresponding transformation of the decay laws at rest has been described via quantum theory and special relativity in term of the model independent MDD \cite{Khalfin,BakamjianPR1961RQT,ExnerPRD1983,HEP_Stef1996,HEP_Shir2004,HEP_Shir2006,TD_StefanovichXivHep2006,UrbPLB2014,GiacosaAPPB2016,GiacosaAPPB2017,UrbAPB2017,GiacosaAHEP2018}. The inverse-power-law decays which are expected over very long times 
are characterized via the threshold of the MDD \cite{BWKnigth1977,FondaGirardiRiminiRPP1978,UrbanowskiEPJD2009,UrbanowskiCEJP2009,SDkelkNovak2010,SDkelkNovak2004,OrgMatNNexp,GEPJD2015}. The transformation of these decays, which is due to the change of reference frame, consists, approximately, in a scaling law \cite{Gxiv2018}. Similar scaling relations are found in the oscillating decay laws of unstable quantum systems which are initially prepared in special superpositions of two, approximately orthogonal, unstable quantum states \cite{HEP_Shir2004}. In fact, if the non-vanishing lower bounds of the two mass spectra coincide, the survival probability $\mathcal{P}_p(t)$ is approximately related to the survival probability at rest by a time dilation. If the lower bounds of the two mass spectra differ, the period of the long-time oscillations of the survival probability transforms according to a factor which is determined by the lower bounds and by the linear momentum of the moving unstable quantum system \cite{GJPA2018}.

In the present research work we have studied the transformation of the decay laws of moving unstable quantum systems over intermediate times. In this regime the survival probability at rest decays exponentially, or according to slower forms, but is faster than the inverse power laws which are expected over very long times. The modulus of the survival amplitude at rest is approximated by a superposition of exponential modes which are obtained from the Prony analysis. The present theoretical construct relies on the following assumptions. Over intermediate times the decay laws are mainly determined by the values of the MDD around the mass of resonance. The contribution of the unphysical negative values of the mass spectrum to the decay law is negligible. The MDD is approximately symmetric with respect to the mass of resonance. These assumptions are essentially based on the property that (symmetric) Lorentzian MDDs provide purely exponential decays of the survival probability at rest \cite{MDDlorentzianGislasonPRA1991,NnExpKumarAJP1992,MDDlorentzianJacobovitsAJP1995,GPoscillatingDecaysQM2012,GiacosaAPPB2016,GiacosaAPPB2017}. In this way, the present approach offers an analytical description of the transformed survival probability in the laboratory reference frame, by performing the Prony analysis of the decay law at rest. This method allows to estimate the time window over which the transformed survival probability decays according to the superposition of the transformed exponential modes. The present theoretical construct describes also the general transformation of the intermediate times which is induced by the change of reference frame.

The relativistic dilation of times is found, approximately, in the transformation of the inverse-power-law decays of moving unstable quantum systems, which are expected over very long times, if the non-vanishing lower bound of the mass spectrum is considered to be the effective mass at rest of the unstable quantum system \cite{Gxiv2018}. According to the present study, the relativistic dilation of times is found, under certain conditions, in the transformation of the decay laws over intermediate times if the mass of resonance is considered to be the effective mass at rest of the moving unstable quantum system. Interpreting experimental works via the present study is beyond the purposes of this paper. However, we believe that the present theoretical construct can be applied to the Prony analysis of detected decay laws at rest. The transformed decay laws, which are obtained via the present approach, can be compared with the detected decay laws of moving unstable systems. This confrontation might provide further insight about the form of the MDD and, more generally, about the description of the decay laws via quantum theory and special relativity.

\appendix\label{A}
\section{Details}

Starting from expression (\ref{A0Int}) of the survival amplitude at rest, the symmetry property, Eq. (\ref{symmMDD}), of the MDD with respect to the mass of resonance $M$ provides the following integral relation,
\begin{eqnarray}
l(t)\sqrt{\mathcal{P}_0(t)}= 
2\int_0^{\infty}\omega\left(M+m^{\prime}\right)
 \cos\left(m^{\prime}t\right) d m^{\prime}, \label{P0MMDint1}
\end{eqnarray}
where $l(t)=\exp \left(\imath \left(\arg\left\{A_0(t)\right\}+M t\right)\right)$. 
The symmetry property induces the right side of Eq. (\ref{P0MMDint1}) to be real-valued. Consequently, the possible values of the function $l(t)$ are uniquely $1$ and $(-1)$. The function $\omega\left(M+m^{\prime}\right)$ is nonnegative and summable, with respect to the variable $m^{\prime}$, over the interval $\left[\right. 0,\infty\left.\right)$ . Therefore, the left side of Eq. (\ref{P0MMDint1}) is continuous. Since the function $\sqrt{\mathcal{P}_0(t)}$ is continuous, the function $l(t)$ is constant in time, $l(t)=l_0$, and is equal to the value $1$ or $(-1)$, uniquely, $l_0=\pm 1$. The MDD is evaluated in terms of the modulus of the survival amplitude at rest, $\sqrt{\mathcal{P}_0(t)}$, by performing the inverse cosine transform of both the sides of Eq. (\ref{P0MMDint1}),
\begin{eqnarray}
\omega\left(M+m^{\prime}\right)=\frac{1}{\pi}\left|
\int_0^{\infty}\sqrt{\mathcal{P}_0\left(t^{\prime}\right)}
 \cos\left(m^{\prime}t^{\prime}\right) d t^{\prime}\right|
. \label{MMDP0int1}
\end{eqnarray}
In this way, the following form is obtained for the transformed survival amplitude,
\begin{eqnarray}
&&\hspace{-3.0em}A_p(t)=\frac{l_0}{\pi}\int_0^{\infty}
\exp\left(-\imath \sqrt{p^2+m^2}t \right) dm \nonumber \\ &&\hspace{0.4em}
\times \int_0^{\infty}
\sqrt{\mathcal{P}_0\left(t^{\prime}\right)}
\Big(\cos\left(\left(m-M\right)t^{\prime}\right)
+\cos\left(\left(m+M\right)t^{\prime}\right)
\Big) dt^{\prime}. \label{ApintccMpMm}
\end{eqnarray}
If the modulus of the survival amplitude at rest is given by Eq. (\ref{A0ExpSum}), the corresponding MDD is evaluated via Eq. (\ref{MMDP0int1}) and results in sum of Lorentzian functions, 
\begin{eqnarray}
\omega\left(m\right)=\sum_{j=1}^N\frac{w_j\Gamma_j/\left(2 \pi\right)}{\left(m-M\right)^2+\Gamma_j^2/4}, 
\label{MDDlorentz}
\end{eqnarray}
which are defined over the whole real line.  Also, Eqs. (\ref{A0ExpSum}) and (\ref{ApintccMpMm}) lead to the following expression of the transformed survival amplitude,
\begin{eqnarray}
&&\hspace{-5.0em}A_p(t)=\sum_{j=1}^N
\frac{ w_j \Gamma_j}{2\pi l_0} 
\int_0^{\infty} \left(\frac{\exp\left(-\imath \sqrt{p^2+m^2}t\right)}{\left(m-M\right)^2+\Gamma_j^2 /4}
+\frac{\exp\left(-\imath \sqrt{p^2+m^2}t\right)}{\left(m+M\right)^2+\Gamma_j^2/4}\right)dm. 
 \label{ApLorentzSum}
\end{eqnarray}
This integral form provides Eq. (\ref{PptHYJ}). The involved integrals are analyzed in Ref. \cite{HEP_Shir2004}. In this regard, some details are reported in the third paragraph of Section \ref{2} via Eqs. (\ref{ApBWMDD}) and (\ref{ApBWMDDasympt}). 

The form (\ref{PptHYJ}) of the survival probability $\mathcal{P}_p(t)$ is approximated via exponential and inverse power laws in Section \ref{31}. Relation (\ref{PpExp1}) is obtained by evaluating the function 
$\Upsilon\left(M,\Gamma_j,p\right)$, via Eqs. (\ref{Upsilon}) 
and (\ref{Lambdamp}), for $\Gamma_N/M\ll 1$. Relation (\ref{XipPl1}) 
is obtained by approximating the function $\Xi\left(M,p,t\right)$, given by 
Eq. (\ref{Xi}), for $pt \gg 1$. Refer to 
\cite{NISThandbook,GradRyzhandBook,AbrSteg} for the definition and the asymptotic expansion of the Bessel Function $J_1\left(pt\right)$, the 
modified Bessel function $Y_1\left(pt\right)$ and the Struve function $\mathbf{H}_1\left(pt\right)$.

The exponential times are estimated by defining the function $K\left(\zeta\right)$ and the variables $\zeta_1, \ldots, \zeta_N$, as below,
\begin{eqnarray}
\hspace{0em}K\left(\zeta\right)=\sqrt{\zeta}\exp\left(-\zeta\right),\hspace{1em}\zeta_j= \frac{\Gamma_j t}{2 \gamma}, 
\label{KZCdef}
\end{eqnarray}
for every $j=1,\ldots,N$. The $j$th exponential mode, which appears in the right side of Eq. (\ref{PpExp1}), dominates over the inverse-power-law term if the following constraint is fulfilled,
\begin{eqnarray}
w_j \exp\left(- \frac{\Gamma_j t}{2\gamma}\right) 
\gg  \frac{p \sum_{i=1}^N
w_i\Gamma_i}{\pi M^2}\sqrt{ \frac{\pi}{2 p t}}.
\label{ExpDomIpl1}
\end{eqnarray}
This constraint is equivalent to the relation below,
\begin{eqnarray}
K\left(\zeta_{j}\right)\gg \xi_{j}.   \label{cnstrEDKl}
\end{eqnarray}
The parameters $\xi_1, \ldots, \xi_N$ are defined by Eq. (\ref{xij}). Inequality (\ref{cnstrEDKl}) holds, by definition, uniquely for the indexes $j=j_1,\ldots,j_{n_0}$, among the indexes $j=j_1,\ldots,j_{N}$. For $\zeta=0$ the function $K\left(\zeta\right)$ vanishes, $K\left(0\right)=0$. The function is positive, $K\left(\zeta\right)>0$, for every $\zeta>0$ and vanishes in the limit $\zeta\to \infty$, $K\left(\infty\right)=0$. The function has one maximum, $K\left(1/2\right)\simeq 0.4289$. Let $\zeta_{ \rm min}$ and $\zeta_{\rm max}$ be the values of the variable $\zeta$ which fulfill the equality $K\left(\zeta\right)= 10^{-2}$. We find $K\left(\zeta_{\rm min}\right)= K\left(\zeta_{\rm max}\right)=10^{-2}$, where $\zeta_{\rm min}\simeq 0.0001$ and $\zeta_{\rm max}\simeq5.4533$. For $\zeta \in \left[\zeta_{\rm min},\zeta_{\rm max}\right]$ the values of the function $K\left(\zeta\right)$ belong to the interval $\left[K\left(\zeta_{\rm min}\right),K\left(1/2\right)\right]$ which is approximately the interval $\left[0.01, 0.43\right]$. Over this interval the minimum order of magnitude of the function $K\left(\zeta\right)$ is fixed to the value $\left(-2\right)$. Therefore, condition, $\xi_j \ll 10^{-2}$, 
guarantees that the constraints (\ref{ExpDomIpl1}) and (\ref{cnstrEDKl}) hold. Consequently, in the laboratory frame $\mathfrak{S}_p$ the modulus of the survival amplitude is properly approximated by the special sum $\sum_{l} {}^{'}$, which appears in Eq. (\ref{ExpDecay1n0}), over the exponential times, which are described via Eqs. (\ref{IpI0l})-(\ref{ExpTn0}).

The transformation of times which is due to the change of reference frame is given by the function $\varphi_p(t)$ which is defined by Eq. (\ref{PpP0def}), or, equivalently, by Eq. (\ref{phipP0Pp}).
 This function exists as the the survival probability 
$\mathcal{P}_0(t)$ is, canonically, a monotonic function of 
time and is, therefore, invertible. The inverse function, $\mathcal{P}^{-1}_0: \left(0\right.,1\left.\right]\to \left[0\right.,+\infty\left.\right)$, can be evaluated numerically, or from the Prony analysis, Eq. (\ref{A0ExpSum}), of the modulus of the survival amplitude at rest, 
\begin{eqnarray}
\mathcal{P}^{-1}_0\left(r\right)=-\frac{2}{\Gamma_1} \ln \left(u\left(r\right)\right). \label{P0invr}
\end{eqnarray}
By performing the change of 
variable $r= \exp \left(-\Gamma_1 t/2\right)$, the function $u\left(r\right)$ is defined for every $r\in \left(0\right.,1\left. \right]$ as the solution of the following equation, 
\begin{eqnarray}
\sum_{j=1}^N w_j u^{\Gamma_j/\Gamma_1}=\sqrt{r}. \label{EqXir}
\end{eqnarray}
The solution is required to be real-valued and belong to the interval $\left(0\right.,1\left. \right]$. The solution which fulfills these properties is unique. Consequently, the function $\mathcal{P}^{-1}_0(r)$ is properly defined over the domain $\left(0\right.,1\left. \right]$. Expression (\ref{PptHYJ}) of the survival probability 
$\mathcal{P}_p(t)$ leads to the form (\ref{phipP0PpHYJ}) of the
 function $\varphi_p(t)$. The linear growth (\ref{varphilinear}) is obtained from the scaling law (\ref{PpPprel}). This concludes the demonstration of the present results.

\end{document}